\documentclass{iopjournal}
\usepackage{amssymb}
\usepackage{amsmath}

\newcommand \MZ [1]{\bgroup\noindent[\textcolor{blue}{\textbf{MZ}: #1}]\egroup\ignorespacesafterend}

\newcommand \PM [1]{\bgroup\noindent[\textcolor{red}{\textbf{PM}: #1}]\egroup\ignorespacesafterend}

\begin{document}

\articletype{Paper} 

\title{Patterns of load, elastic energy and damage in network models of architected composite materials}

\author{Christian Greff$^{1,2}$, 
Leon Pyka$^{1,2}$, 
Michael Zaiser$^1$\orcid{0000-0001-7695-0350} and
Paolo Moretti$^{1,*}$\orcid{0000-0002-2936-1403}}

\affil{$^1$Dept. of Materials Science, WW8-Materials Simulation, FAU Universit\"at Erlangen-N\"urnberg, Dr.-Mack-Stra{\ss}e 77, 90762 F\"urth, Germany}
\affil{$^2$ FAU Competence Center Scientific Computing, Martensstr. 5a, 91058 Erlangen, Germany}
\affil{$^*$Author to whom any correspondence should be addressed.}

\email{paolo.moretti@fau.de}


\begin{abstract}
We investigate the role of architected thin films in the interfacial failure properties of bi-layer composites. Our results show that, while graded structures can be used to prescribe failure at the interface, they do not offer significant advantages in terms of fracture toughness. Hierarchically patterned layers can localize failure at the interface and simultaneously enhance interface toughness, by enforcing a buffer region where elastic energy is dissipated in the form of diffuse damage, so that no stress concentration can drive crack growth. To analyze these mechanisms, the associated patterns of local load redistribution and the soft deformation modes, we develop a network formalism that brings together concepts of discrete differential geometry and spectral graph theory.

\end{abstract}

\section{Introduction}
Fracture and failure can be envisaged as processes that brittle and semi-brittle materials initiate to cope with excessive load. Essentially, the stored potential energy is dissipated by micro-crack formation. While in homogeneous materials this mechanism triggers rapid crack expansion and catastrophic failure, heterogeneity in structure and local constitutive behavior may slow the process down, into a two-step failure mode where an extended crack nucleation stage precedes crack propagation \cite{Bonamy2006_PRL,alava2006statistical,alava2008role}.

In recent years, research on fracture of architected materials has focused on the enhanced mechanical properties that derive from complex microstrucural design. For instance, hierarchically architected microstructures have been explored as an approach to impede crack propagation by inducing widespread crack deflection and arrest \cite{Lakes1993,fratzl2007nature,gao2006application}, a novel type of failure mode that often results in higher fracture toughness and insensitivity to pre-existing flaws \cite{Zaiser2022,pournajar2023failure} and which has been recently associated with the formation of multifractal fracture surfaces \cite{Hosseini2021,hosseini2023enhanced,Greff2024_SciRep}.  In hierarchically architected materials, elastic load transmission in a hard and brittle matrix is modulated by the presence of  soft and compliant lamellar inclusions or void-like planar gaps \cite{Sen2011,mirzaeifar2015defect,Moretti2018}, which interrupt stress transmission and facilitate crack deflection, promoting diffuse damage rather than the propagation of critical flaws \cite{Sun2012, Jiao2015, Gao2006, Rho1998, Gautieri2011, Sen2011,roemer2008prion,lu2023silk}.

The mechanical advantages of hierarchical microstructures have also been studied in the context of thin-film interface failure \cite{Puglisi2013_PRE,Esfandiary2022,Greff2024_SciRep} and friction \cite{Costagliola2016_PRE,Costagliola2022_IJSS}, often inspired by the Gecko foot as a paradigm of smart, recyclable adhesives \cite{KimJAST_2007,Sauer2014_JAST,Bhushan2009_PTRS,Gao2005_MM}. In analogy with the case of bulk fracture, numerical models of failure of interfaces that can be modelled as hierarchically patterned network structures confirm the appearance of crack arrest phenomena. In these systems, the soft lamellar inclusions often take the form of planar voids or cuts emanating from the interface in such a manner that the system becomes more sparse and more compliant in proximity of the interface. This type of arrangement offers the added benefit of promoting damage accumulation near the interface, allowing for tuning the failure location. We notice here that what prescribes failure location in these systems likely is the \textit{graded} nature of the density/sparsity patters. Indeed, research on graded microstructures focuses on the possibility of tuning materials behavior by imposing simple design constraints, such as the variation of matter density in precise directions relative to the expected external loads \cite{bai2020_ijms,niknam2020_md}. Contrary to hierarchical structures, generic graded structures may implement variation in sparsity by randomly distributing micro-voids, rather than multi-scale extended defects and for this reason it is reasonable to expect failure to advance according to the standard nucleation-and-propagation paradigm of heterogeneous materials, rather than the crack arrest scenario described above. However, no fair comparison between the fracture properties of graded structures and their more complex hierarchical counterparts is found in the literature. 

Interface failure cannot be completely understood without considering the role played by the substrate: architected material models of this type are composite models in nature, there can be no top layer without a substrate. While studies of this type have often focused on simplified geometries with infinitely stiff substrates, this choice may often be dictated by numerical convenience \cite{Esfandiary2022,Greff2024_SciRep}. A proper substrate, modeled for instance as a heterogeneous, non-architected layer, may promote patterns of stress redistribution and damage propagation  that compete with the response of the architechted top layer, possibly nullifying its functional and mechanical advantages.  

Here we model interface failure of architected thin films deposited on compliant and heterogeneous substrates. Our structures are subjected to film-normal, uniaxial tensile loads, thus mimicking peeling phenomena. We consider both hierarchical and comparable graded film structures for the top layer, and allow the substrate to have varying stiffness/hardness, in order to assess the role played by stress redistribution and damage accumulation in both the top layer and the substrate. The local mechanical response is simulated here using the Random Fuse Model (RFM) where the material is modeled as a network of load carrying edge-elements obeying scalar constitutive laws \cite{deArcangelis1985_JPL}. While more sophisticated approaches exist, where a full description of elements  as Timoshenko beams is used \cite{Hosseini2021,hosseini2023enhanced}, simpler scalar models such as the RFM become relevant in large-scale statistical studies like ours where the computational cost of beam simulations might be significant. 

An advantage of the RFM is that the materials microstructure is broken down to a fuse network, and more specifically the elastic behavior can be mapped to the steady DC response of a resistor network, subject to an eternally applied voltage drop. The statistical physics community has resorted to resistor network simulations to explore the emergence of scaling in fracture and failure in heterogeneous materials \cite{alava2006statistical,alava2008role,Picallo2009_PRL,Barai2013_PRE}. In more recent years related concepts of network theory \cite{Berthier2019_PNAS,Pournajar2022,Konrad2022_Polymers}, spectral graph theory \cite{Moretti2019_EPJB,Moretti2019_PNAS} and knot theory \cite{Puhlmann2025_Polymers} have been invoked to predict more generic aspects of the mechanical behavior of materials with inherently complex microstructures.

Our simulation results show that while both hierarchical and graded structures allow for localization of cracks at the interface, the greater complexity of hierarchical systems yields significant gains in terms of work of fracture, making such systems inherently more resilient. We develop a network theoretical approach that allows us to spatially resolve local constitutive laws and patterns of stress/energy storage, using the concept of discrete $k-$forms. We find that the unique behavior of hierarchical systems can be attributed to specific patterns of stress redistribution that create a \textit{buffer region} near the interface, where potential energy is dissipated in the form of diffuse damage and cracks do not propagate. We propose the use of local spectral indicators such as the local density of states to identify this buffer region.  

\section{Numerical model}

\subsection{Network model of a quasi-brittle bi-layer material}

\begin{figure}[h]
\centering
\includegraphics[width=0.9\linewidth]{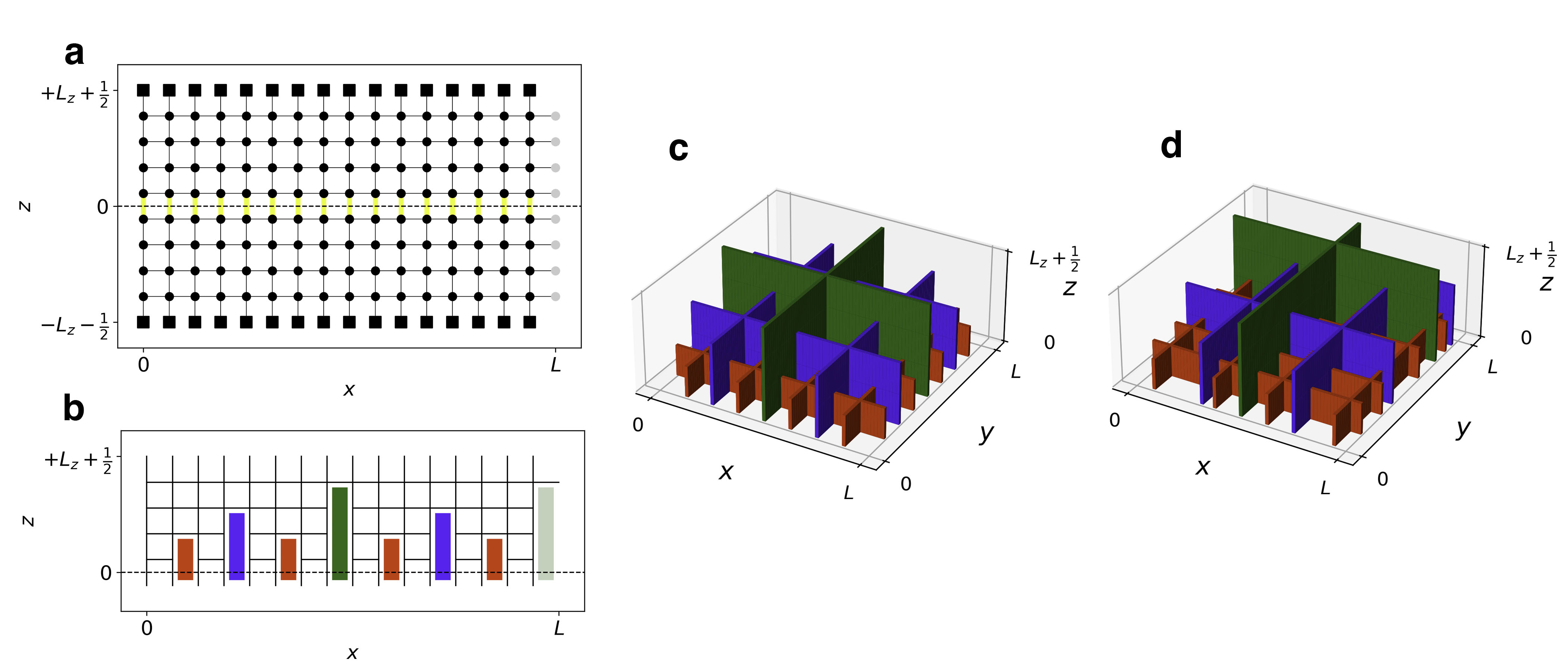}
\caption{Network model of a bi-layer composite, and details of the construction of a hierarchical H top layer. Figure partially adapted from \cite{Greff2024_SciRep}. The case of $s=4$ is shown.  (a) Nodes and edges are distributed in a 3D cubic lattice. The boundaries at $x=L$ (shown in gray) and $y=L$ (not shown) are periodic. The boundaries at $z=\pm(L_z+1/2)$ (square nodes) are not periodic. Edges at the interface ($z=0$) are plotted in yellow;  (b) In order to generate a hierarchical H top layer, in the region with $z>0$, $x/y$-edges are recursively removed to form cuts.  The case of a \textit{deterministic} hierarchical cut pattern is shown. Cuts of levels $h=1$, $h=2$, $h=3$ are shown in red, blue, green respectively. Periodic boundaries  are introduced as additional highest-level cuts (pale green). (c) 3D view of the deterministic cut structure of (b). (d) 3D view of the final H cut pattern, obtained by shuffling cut positions.}
\label{fig:networks}
\end{figure}

We model a bi-layer material as a network of nodes connected by load-carrying edges of length $\ell = 1$. Construction of the bi-layer structure proceeds as follows: First, nodes are placed on the sites of a cubic primitive lattice, and adjacent nodes are connected by edges of unit length to form a perfect simple cubic lattice of size $L\times L\times (2 L_z + 1)$, with periodic boundary conditions in the $x$ and $y$ directions (Figure \ref{fig:networks}a).

Dimensions are chosen such that, given a positive integer $s$, $L=2^s$ and $L_z=s$. The total number of nodes in the system is $N = L^2\times 2(L_z+1)$. Each node $i$ is associated with three discrete coordinate values $[x,y,z]$. The coordinates in the $xy$ plane are integers $x,y = 1,2,\dots,L$ and the coordinates in the $z$ direction are half integers $z = -L_z-1/2,\dots, L_z + 1/2$. The plane $z=0$ defines an interface (yellow edges in Figure \ref{fig:networks}a) between a substrate layer S for which all nodal $z$ coordinates are negative, and a top layer T with positive $z$. The height of each layer is $L_z$. The total height of the system, comprising the two layers and the interface, is thus $L_z'=2L_z+1$

Starting from this layout, the top and bottom layers are endowed with specific, architected microstructure by selective removal of edges. While all edges oriented along the $z$ direction ($z$-edges in the following) are retained, edges along $x$ and $y$ directions ($x/y$-edges in the following) are selectively removed, following a removal pattern that depends on whether the layer is of the hierarchical H, graded G, or random R type.

In the H case, a hierarchical edge removal pattern follows the rules introduced in References \cite{Esfandiary2022,Greff2024_SciRep} i.e. $x/y$-edges are selectively removed depending on the distance from the interface in an $(s-1)$-step process, in order to mimic hierarchical contact for a system comprising $s$ hierarchical levels, see Figure \ref{fig:networks}b-d). Edge removal in the H system introduces planar cuts (or gaps) parallel to either the $xz$ or the $yz$ planes and spanning the whole system in the $x$ or $y$ directions. The cuts are introduced by removing parallel $x/y$-edges with the same value of either $x$ or $y$, and with distances from the interface in the range between $z=1/2$ and $z=h-1/2$, where $1 \le h < s$ is the height of the cut. For $1\le h<s$, the $h$-th hierarchical level of the cut pattern comprises $2^{s-h-1}$ cuts of height $h$ in the $x$ direction and an equal number in the $y$ direction. This means that, in a layer with distance $h > 0$ above the interface, a total of $2^{2s+1-h}$ $x/y$-edges out of the $2^{2s+1}$ available ones are removed. The density of missing edges is maximum near the bottom boundary, where it reaches $\rho_{\rm g} = 1/2$, and then decreases exponentially with increasing distance from the interface as $\rho_{\rm g} = 2^{-h}$. In our study, we randomly assign (or shuffle) the positions of the cuts (Figure \ref{fig:networks}d), thus producing a different stochastic variant of the pattern in every system realization. 

\begin{figure}[t]
\centering
\includegraphics[width=0.9\linewidth]{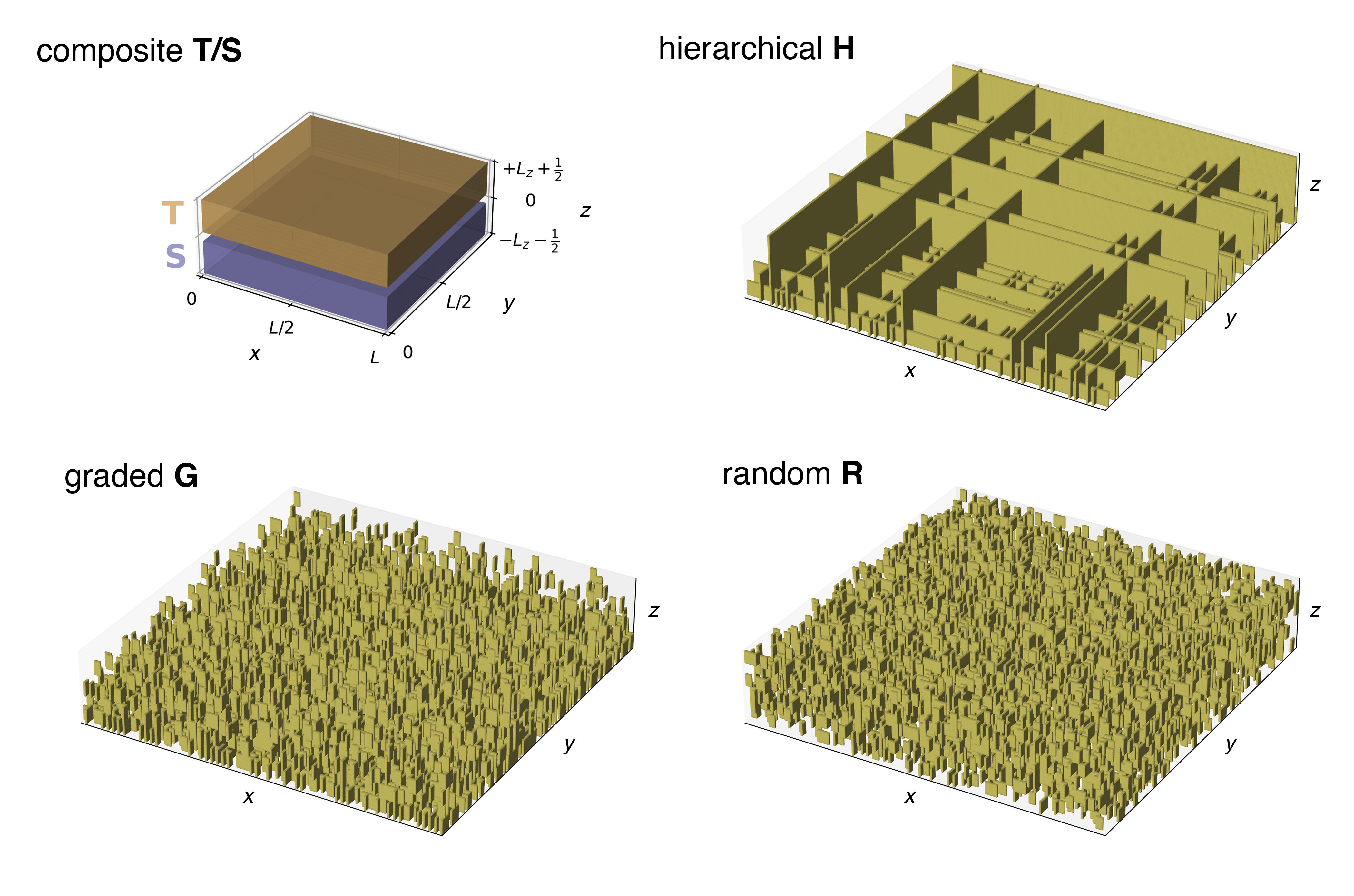}
\caption{Schematic representation of the T/S bi-layer composite model. The top layer T can be of type H (hierarchical), G (graded) and R (random). The substrate layer S is of type R. Microstructures of each layer type are plotted by displaying the portions of the layer that are removed. H layers exhibit patterns of extended cuts of different heights, originating from the bottom boundary and ensuring that the layer is sparser near that boundary. G layers exhibit the same density profile as H, but matter is removed in the form of randomly distributed voids. R layers too exhibit randomly distributed voids, but the density is homogeneous. The globally averaged sparsity/density is the same for all systems.    }
\label{fig:model}
\end{figure}

In the graded G case, the density of removed $x/y$-edges exhibits the same height dependency as in the H case, with a fraction $2^{-h}$ of all $x/y$-edges removed in a layer that has distance $h-1/2$ from the interface, however, the removed edges are selected at random for each $h$. Finally, in the R case, the total number of removed $x/y$-edges is retained, and no constraints are imposed based on height. These construction criteria ensure that the H and G systems share the same $x/y$-edge number per layer, while the R system only retains the global $x/y$-edge number. 

In the T/S composite (Figure \ref{fig:model}), the top T network (of type H, G, or R) is connected to a bottom S network of type R by the set of parallel $z$-edges with midpoints at $z=0$, i.e. the interface is represented by a layer of vertical edges (in yellow in Figure \ref{fig:networks}a). As crack surfaces will be identified as bundles of vertical edges, cracks at the interface will thus have the location $z=0$. We consider two types of structures: Structures where the interface is initially intact (i.e., which have $L \times L$ interface edges in the plane $z=0$), and structures that contain a preexisting crack of size $L\times a$ in the $z=0$ plane.  

The $N$ nodes of the final T/S model are labeled using indices $i=1,2,\dots,N$, and the connectivity information is stored in the adjacency matrix $\mathsf{A}$, whose generic element $A_{ij}$ is $1$ if nodes $i$ and $j$ are  connected by an edge $e$, and $0$ otherwise. The $\tilde{N}$ edges are also labeled as integers, using indices $e=1,2,\dots,\tilde{N}$. We call $\Omega$ the set of all nodes, and $\partial\Omega$ its boundary, consisting of all nodes at $z=-L_z - 1/2$ and $z=L_z + 1/2$. The set of all non-boundary nodes is $\Omega\setminus\partial\Omega$.

\subsection{Elasticity}
To model interface failure, we consider the simple case in which the bottom boundary ($z=-L_z - 1/2$) is fixed, while a displacement is imposed at the top boundary ($z=L_z + 1/2$). We model the elastic response of the system and its failure in the RFM framework. Each node $i$ has a displacement-like variable $u_i$ and each edge $e$ carries a force-like variable $f_{e}$. Since elastic variables are envisaged as scalars, the elastic behavior of a generic edge $e$ connecting nodes $i$ and $j$ is described by a scalar version of Hooke's law of the form 
\begin{equation}
\frac{f_{e}}{X}=E_{e} \frac{u_i-u_j}{\ell}
\end{equation}
 where $X$ and $\ell$ are the edge cross-section and length, respectively, and the elastic modulus $E_{e}$ is the proportionality constant between edge stress $f_{e}/X$ and edge strain $(u_i-u_j)/\ell$. We use $\ell$ as the unit of length and, for convenience of notation, we introduce the edge stiffness $\kappa_{e} = E_{e}X/\ell$ in a form that is reminiscent of axial stiffness in beam elasticity.
 We assume that no external body forces act on the system, so that elastic equilibrium at each non-boundary node $i\in\Omega\setminus\partial\Omega$ is imposed through the system of algebraic equations
\begin{equation}\label{eq:discrete_laplace}
\sum_{j\in\Omega}L_{ij}u_j=0 \;\;\;\textrm{with} \;\;\;i\in\Omega\setminus\partial\Omega
\end{equation}
where the displacements $u_i$ at non-boundary nodes $i\in\Omega\setminus \partial\Omega$ are the unknowns, the displacements $u_i$ at boundary nodes $i\in\partial\Omega$ are fixed, and $L_{ij}$ are the elements of the discrete Laplace operator $\mathsf{L}$. Introducing the weight matrix with the generic element $W_{ij}= \kappa_e A_{ij}$, and labelling $e$ the edge that connects nodes $i$ and $j$, one has  $L_{ij}=\delta_{ij}\sum_{l\in \Omega} W_{jl}-W_{ij}$ (weighted graph Laplacian matrix). $\mathsf{L}$ is the network equivalent of the heterogeneous Laplace operator $-\nabla \kappa(\mathbf{r})\nabla$ in the continuum. Later on (Equation \ref{eq:laplace_C}) we will introduce an equivalent formula for $\mathsf{L}$, which more closely resembles its continuum counterpart.

Solutions of Equation \ref{eq:discrete_laplace} depend on the choice of boundary conditions. 
External loads in our uni-axial geometry are applied in the form of displacements, e.g. fixing $u_i=0$ at the lower boundary and $u_i=U>0$ at the upper boundary \cite{Greff2024_SciRep}. If $r$ is the number of non-boundary nodes, the resulting system of algebraic equations has $r$ degrees of freedom and is solved  using a parallel sparse direct solver, available at our supercomputing infrastructure.

\subsection{Heterogeneity in constitutive behavior}

\begin{figure}[tbhp]
\centering
\includegraphics[width=0.7\linewidth]{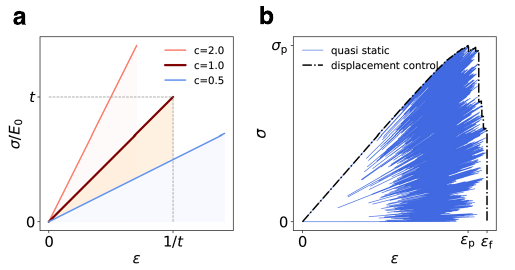}
\caption{(a) Average constitutive behavior of the edges in the substrate layer, for different values of the parameter $c$. The average failure threshold is chosen such that curves of different stiffness have, on average, the same area (work of failure). (b) Typical stress-strain curve under displacement control (thick dot-dashed black line), constructed by enveloping the stress-strain curve obtained from a quasi-static simulation (thin solid blue line).}
\label{fig:details}
\end{figure}

On the edge level we consider ideal elastic-brittle behavior of the individual edges $e$. Thus, each edge $e$ behaves elastically until the force $f_{e}$ that it carries reaches a force threshold $t_{e}$, and is irreversibly removed as soon $f_{e}>t_{e}$ \cite{alava2006statistical}. Each network realization is characterized by different sets of $t_{e}$, extracted from a Weibull distribution with mean value $\bar{t}$ and shape parameter $k=4$ \cite{Greff2024_SciRep}. 
In order to model layers of different strength and/or compliance, we vary both $E_{e}$ and  $\bar{t}$ in such a way that the average area under the corresponding edge-wise stress-strain curve (average work of edge failure) is conserved (Figure \ref{fig:details}a). Specifically, for all edges $e$ in a layer, we define a common tuning parameter $c$ such that  $\kappa_{e} = c \kappa_0 $, $E_{e} = c E_{0} $ and $\bar{t}_{e}=c^{-1}\bar{t}_0$. High values of $c$ define a stiff material with low failure strain, whereas low values of $c$ define a soft material with high failure strain. In our simulations we choose a fixed value $c=1$ for the top T layer and also for the edges at $z=0$ which define the interface, but we allow for variations of $c$ in the substrate S layer where we consider three scenarios: (a) $c=1$, i.e. T and S layers share the same elastic response; (b) $c=2$ i.e. the S substrate is stiffer than the T layer; (c) $c=0.5$, i.e. the S substrate is softer than the T layer. In all simulations, we measure forces in units of $t_0$ and stresses in units of $\sigma_0=t_0/X$. 

\subsection{Simulation protocol}

The global stress is computed from the global force $F$ acting on a cross section of the system as $\sigma = F/L^2$. Similarly, the global strain is computed starting from the boundary displacement $U$ as $\epsilon = U/L'_z$;  it is measured in units of $\sigma_0/E_0$. We perform displacement-controlled simulations, using the standard quasi-static simulation procedure, where a constant displacement $U$ is applied and no body forces are present \cite{alava2006statistical}. At every step, the equilibrium equations are solved for every non-boundary node $i$, edge forces are computed, the edge with the maximum load factor $\rho_{\rm max}:={\rm max}_{e}(f_{e}/t_{e})$ is identified as the weakest and removed. The local and global forces and displacements are rescaled by the factor $\rho_{\rm max}$, thus effectively setting the displacement to the precise value when the weakest edge fails. The simulation is terminated when the stress (or, equivalently, the global elastic modulus of the system) reaches zero, indicating complete failure. In the failed state, the displacement field exhibits discontinuities in such a manner that, for each column of nodes with common $[x,y]$, the displacement is zero for $z \le z_{\rm f}(x,y) - 1/2$, and $U$ for $z = z_{\rm f}(x,y) + 1/2$. The loci of the displacement discontinuity at $z = z_{\rm f}(x,y)$ always correspond to a system-spanning cluster of broken edges and define the final fracture surface. 

While quasi-static simulation method describes an idealized protocol, where at every step displacements are adjusted to the minimum values allowing for the removal of the weakest link, the desired protocol in which displacement is monotonically increased can be obtained by enveloping the resulting stress-strain curve. This procedure is exemplified in Figure \ref{fig:details}b. 

For an individual run, the peak stress $\sigma_\mathrm{p}$ is the maximum value of $\sigma$. The work of failure $W_\mathrm{f}$ (per unit volume) is the area under the stress-strain curve, and  quantifies the energy per unit volume that is necessary to reach failure, including the post-peak regime. For specimens containing an initial interface notch of width $a$, the specific work of failure is defined as $w_\mathrm{f}=(1-a/L)^{-1}W_\mathrm{f}$, i.e. the area under the stress-strain curve normalized by the initially intact surface area fraction $(1-a/L)$.

In our statistical study, global fracture strength is evaluated by measuring the peak stress $\sigma_\mathrm{p}$ and the specific work of failure $w_\mathrm{f}$, for both notched and un-notched systems.   Both $\sigma_\mathrm{p}$ and $w_\mathrm{f}$ are averaged over multiple network realizations. We consider systems of sizes $L=128$, corresponding to a film thickness $L'_z=2L_z+1=
15$ (equivalent to $s=7$ hierarchical levels in H layers). Un-notched samples ($a=0$) of this size contain  $\tilde{N} = 6.4\times 10^5$ edges each. During each run, approximately $3-4 \times 10^4$ edges are broken before failure. Prior to each edge breaking event, the equilibrium problem is solved and then the failure criterion applied.

\section{Theory}
Our systems are characterized by spatial heterogeneity of both structure and constitutive behavior. In order to understand a system of this type, one needs a theory that correctly resolves quantities in space. Importantly, certain quantities such as displacements are evaluated at nodes, and are thus discrete 0-forms. Other quantities such as forces are evaluated at edges, and thus can be interpreted as discrete 1-forms. This distinction is inherent to discrete systems and discretization approaches to exterior calculus \cite{Hirani2003_Thesis,Desbrun2008_book,Grady2010_Book,Crane2017_Glimpse,Arnold2018_Book}, but somewhat underdeveloped in the RFM literature. In what follows, we reformulate the scalar elastic problem in terms of $k$-forms, addressing this conceptual gap, and adopting for the most part the formalism of Forman \cite{Forman1993_Laplacians}. Later we move to the spectral analysis of our model, providing spatially resolved indicators that can be used as failure predictors. The modelling tools introduced in this section will be used in the analysis of simulation results.

\subsection{Material model, Laplace operator and elastic energy}
We recall that our system is composed of $N$ nodes, labelled $i=1,2,\dots,N$ and $\tilde{N}$ edges, labelled $e=1,2,\dots,\tilde{N}$. We express the displacements as $\mathbf{u}$, the $N$-dimensional column vector of elements $u_i$. Similarly, we express the forces as $\mathbf{f}$, the $\tilde{N}$-dimensional column vector of elements $f_e$. In order to measure forces on edges we need to assign an \text{intrinsic} orientation to every edge. To this end, we introduce the signed incidence matrix $\mathsf{M}$, i.e. the rectangular $N\times \tilde{N}$ matrix with elements that fulfill 
\begin{equation}
m_{ie}=
\begin{cases}
      -1 & \text{if $e$ leaves $i$}\\
      1 & \text{if $e$ arrives at $i$}\\
      0 & \text{if $e$ is not connected to $i$}
\end{cases} 
\end{equation}
For a generic network with non-directed edges, the assignment of ``incoming'' and ``outgoing'' edges is arbitrary but subject to the condition that, for an edge $e$ connecting nodes $i$ and $j$, the relation $m_{ie}m_{je} = -1$ must always be fulfilled. Here, however, we are dealing with a network implemented on a three-dimensional Euclidean space where each node is located on the sites of a primitive cubic lattice and each edge is aligned either with the $x$, $y$ or $z$ coordinate axis of a Cartesian coordinate system. Each node $i$ is associated with spatial coordinates $ [x_i,y_i,z_i]$. Then, for two nodes connected by an edge $e$ we use the sign convention  $m_{ie} = {\rm sign}(x_i + y_i + z_i - x_j - y_j - z_j) = -m_{je}$. In simple words this means that an edge that points from node $i$ in positive $x$, $y$ or $z$ direction counts as outgoing, and an edge that points in negative $x$, $y$ or $z$ direction is counted as incoming. Edges crossing the periodic system boundaries are counted as incoming at $x=0$ or $y=0$, and as outgoing at $x=L$ or $y=L$. Using this convention allows us to directly connect the topological structure encoded in the incidence matrix with the spatial structure of the network. 

We also introduce the $\tilde{N}\times \tilde{N}$ diagonal matrix $\mathsf{C}$, whose diagonal element $c_{ee}$ is the stiffness of edge $e$: if edge $e$ connects nodes $i$ and $j$, then $c_{ee}=\kappa_{ij}$. $\mathsf{C}$ is known in the literature as the conductance matrix. 

The displacement $\mathbf{u}$ is a function defined on nodes, i.e. a discrete 0-form or a 0-cochain, while the force $\mathbf{f}$ is a function defined on edges, i.e. a discrete 1-form or 1-cochain. In the following, we will just refer to $k$-forms, always implying \textit{discrete} $k$-forms (as opposed to \textit{differential} $k$-forms), or $k$-cochains.  What makes this classification powerful is the fact that we can map $0$-forms to $1$-forms via the operator $\mathsf{d}$, which is referred to as discrete exterior derivative or coboundary operator. In particular, when applied to $\mathbf{u}$ as $\mathsf{d}\mathbf{u}$ it produces a 1-form which is the discrete differential of the displacement: for edge $e$ connecting nodes $i$ and $j$, the $e$-th element obeys $((\mathsf{d}\mathbf{u}))_e=u_j-u_i$. The sign convention introduced above then allows us to connect the exterior derivative (a network operator) with the spatial derivative. 

$\mathsf{d}$ has a known matrix representation in terms of the incidence matrix introduced above, namely $\mathsf{d}=\mathsf{M}^\mathrm{T}$. Interestingly, both $\mathsf{d}\mathbf{u}$ and $\mathbf{f}$ are $1$-forms, and while they are obviously related via the constitutive law (Hooke's law)
\begin{equation}\label{eq:hookes_d}
\mathbf{f}=\mathsf{C}\mathsf{d}\mathbf{u},
\end{equation}
we can embed the constitutive law in the metrics. To this end, we endow the space of $0$-forms with the inner product
\begin{equation}\label{eq:bilinear_0}
\langle \mathbf{v}, \mathbf{w}\rangle_0
= \mathbf{v}^\mathrm{T}\mathsf{W_0}\mathbf{w},
\end{equation}
where $\mathbf{v}^\mathrm{T}$ is the row vector obtained transposing the column vector $\mathbf{v}$, $\mathsf{W}_0$ is a real $N\times N$ diagonal matrix, $\mathbf{v}^\mathrm{T}\mathsf{W}_0$ is the dual $0$-form of $\mathbf{v}$, and the inner product can be interpreted as $\mathbf{v}^\mathrm{T}\mathsf{W}_0$ acting on the $0$-form $\mathbf{w}$, expressed as a matrix multiplication. Similarly, the inner product for $1$-forms is defined as 
\begin{equation}\label{eq:bilinear_1}
\langle \mathbf{h}, \mathbf{k}\rangle_1
= \mathbf{h}^\mathrm{T}\mathsf{W_1}\mathbf{k},
\end{equation}
$\mathsf{W}_1$ is a real $\tilde{N}\times \tilde{N}$ diagonal matrix, and the same considerations on duality as above hold.
As we deal with a network of resistors, the natural choices of  metrics are $\mathsf{W}_0=\mathsf{I}$, so that $\langle\cdot,\cdot\rangle_0$ is the scalar product $\langle\cdot,\cdot\rangle$ in $\mathbb{R}^N$, and $\mathsf{W}_1=\mathsf{C}$ \cite{Forman1993_Laplacians}.
Thanks to this choice, one can  simply express Hooke's law by saying that $\mathbf{f}^\mathrm{T}$ is the dual 1-form of $\mathsf{d}\mathbf{u}$\footnote{We could write this as  $\mathbf{f}^\mathrm{T}=(\mathsf{d}\mathbf{u})^\flat$. Alternatively, if we were using the index notation of Ricci calculus, we would say that $\mathbf{f}^\mathrm{T}$ is obtained from $\mathsf{d}\mathbf{u}$ by \textit{lowering the indices}.}.

More importantly, our metrics allow us to compute the Laplace operator $\mathsf{L}$ as a Hodge Laplacian for $0$-forms, as 
\begin{equation}\label{eq:hodge}
\mathsf{L} = \mathsf{\delta}\mathsf{d},
\end{equation}
where $\mathsf{\delta}$ is the adjoint (codifferential) of $\mathsf{d}$ w.r.t the inner products above, which satisfies
\begin{equation}\label{eq:duality}
\langle \mathsf{\delta}\mathbf{h},\mathbf{v} \rangle_{0} =  \langle \mathbf{h},\mathsf{d}\mathbf{v} \rangle_{1}.
\end{equation}
Applying the identities in Equations \ref{eq:bilinear_0} and \ref{eq:bilinear_1} to Equations \ref{eq:duality} and \ref{eq:hodge}, we find that in general $\mathsf{\delta}=\mathsf{W}_0^{-1}\mathsf{d}^\mathrm{T}\mathsf{W}_1$, 
$\mathsf{L}=\mathsf{W}_0^{-1}\mathsf{d}^\mathrm{T}\mathsf{W}_1\mathsf{d}$ and more specifically in our scalar-elastic model
\begin{equation}\label{eq:laplace_C}
\mathsf{L}=\mathsf{d}^\mathrm{T}\mathsf{C}\mathsf{d},
\end{equation}
which is known in the literature as the generalized Laplace operator for a resistor network with heterogeneous conductances.

We can now use this formalism to compute and resolve energies. The elastic energy $E$ stored in the system because of the displacements $\mathbf{u}$ is given by the sum of the potential energies stored in every edge, and can be expressed as any of the following equivalent forms
\begin{equation}\label{eq:energy_bilinear}
E = \frac{1}{2}\mathbf{u}^\mathrm{T}\mathsf{L}\mathbf{u} 
= \frac{1}{2}(\mathsf{d}\mathbf{u})^\mathrm{T}\mathsf{C}\mathsf{d}\mathbf{u}
= \frac{1}{2}\mathbf{f}^\mathrm{T}\mathsf{d}\mathbf{u}
=\frac{1}{2}\langle \mathsf{d}\mathbf{u}, \mathsf{d}\mathbf{u} \rangle_1
\end{equation}
When computing the elastic energy $E'$ restricted to a subset of edges, we can introduce the conductance matrix $C'$ where we set the conductance of edges not included in the subset to zero, so that
\begin{equation}\label{eq:energy_restricted}
E'
= \frac{1}{2}(\mathsf{d}\mathbf{u})^\mathrm{T}\mathsf{C}'\mathsf{d}\mathbf{u}
= \frac{1}{2}\mathbf{u}^\mathrm{T}\mathsf{d}^\mathrm{T}\mathsf{C}'\mathsf{d}\mathbf{u},
\end{equation}
which we shall use later in our computational study. Finally, we note that, while the individual components of $\mathsf{d}\mathrm{u}$ and $\mathbf{f}$ depend on the initial choice of edge orientations, the elastic energy does not, since the derivatives are squared. This is equivalent to what we observe in the continuuum, where the components of a force vector depend on the choice of frame of reference, but energies are invariant.

\subsection{Equilibrium problem}
Here we develop the matrix formulation of the equilibrium problem sketched in Equation \ref{eq:discrete_laplace}. To fix the notation, we choose our node labels such that the non-boundary nodes have indices $i=1,2,\dots,r$, and boundary nodes $i=r+1,r+2,\dots,N$. The full system of algebraic equations describing equilibrium, and including boundary nodes, is
\begin{equation}\label{eq:blocks}
\begin{pmatrix}
\mathsf{K} & \mathsf{R} \\
\mathsf{R}^\mathrm{T} & \mathsf{S}
\end{pmatrix}
\begin{pmatrix}
\mathbf{v} \\
\mathbf{w}
\end{pmatrix}
=
\begin{pmatrix}
\mathbf{f}^* \\
\mathbf{g}^*
\end{pmatrix}
\;\;\;\;\; \textrm{with } 
\mathsf{L}=
\begin{pmatrix}
\mathsf{K} & \mathsf{R} \\
\mathsf{R}^\mathrm{T} & \mathsf{S}
\end{pmatrix}
\textrm{ and }
\mathbf{u}= 
\begin{pmatrix}
\mathbf{v} \\
\mathbf{w}
\end{pmatrix},
\end{equation}
where the matrix on the left is the Lapacian $\mathsf{L}$ in block matrix representation, and $\mathsf{K}$ is the $r\times r$ \textit{stiffness} matrix, i.e. the non-singular matrix that enters the actual solution of the algebraic problem after boundary conditions are applied. $\mathsf{L}$ is thus the Laplacian of the \textit{free} problem, and $\mathsf{K}$ the Laplacian of the constrained problem. Accordingly, $\mathbf{v}$ and $\mathbf{f}^*$ are column vectors of size $r$, corresponding to displacements and external forces acting on non-boundary nodes, while $\mathbf{w}$ and $\mathbf{g}^*$ are the column vectors of size $N-r$ corresponding to the same quantities acting on boundary nodes. The advantage of this formulation is that any algebraic problem of this type can be expressed through block matrix operations, which can be efficiently implemented in numerical studies using sparse array libraries. In our case, we solve a Dirichlet boundary problem ($\mathbf{w}$ is fixed, $\mathbf{v}$ is unknown), with no external body forces ($\mathbf{f}^*=0$, resulting in the null right-hand side in Equation \ref{eq:discrete_laplace}). Our problem is thus completely defined by

\begin{align}
\mathsf{K}\mathbf{v} &=-\mathsf{R}\mathbf{w} 
\label{eq:solver_problem}\\
\mathbf{g}^* &= \mathsf{R}^\mathrm{T}\mathbf{v}+\mathsf{S}\mathbf{w}
\label{eq:solver_forces}
\end{align}
where Equation \ref{eq:solver_problem} is the algebraic problem supplied to a numerical solver. Equation \ref{eq:solver_forces} allows us to compute the forces on every boundary node. The sum of the elements of $\mathbf{g}^*$ restricted to the top boundary nodes equals the global force, as well as the negative sum of the elements of $\mathbf{g}^*$ restricted to the bottom boundary nodes.

\subsection{Spectral formalism}
We consider the spectral representation of our Laplace operators $\mathsf{L}$ and $\mathsf{K}$,
\begin{equation}
\mathsf{L}=\sum_{m=1}^{N}\lambda_m \boldsymbol\phi_m \boldsymbol\phi_m^\mathrm{T}\;\;\;\;\;\;
\mathsf{K}=\sum_{m=1}^{r}\mu_m \boldsymbol\psi_m \boldsymbol\psi_m^\mathrm{T}
\end{equation}
by means of their $m$-th eigenvalues and corresponding \textit{unit} eigenvectors. As both operators are real and symmetric, eigenvalues and eigenvectors are real. Eigenvalues are indexed in ascending order. $\mathsf{L}$ is positive semi-definite and has as many null eigenvalues as the number of connected components in the network, while, if the boundary value problem is well posed, $\mathsf{K}$ is positive definite, so that for a network with a single connected component
\begin{equation}
0=\lambda_0 < \lambda_1 \le \lambda_2 \le \dots \le \lambda_N 
\;\;\;\;\;\;\
0<\mu_1 \le \mu_2 \le \dots \le \mu_r. 
\end{equation}
Importantly, while the eigenvalue spectra of $\mathsf{L}$ and $\mathsf{K}$ are not the same, they interlace, i.e. every eigenvalue of $\mathsf{K}$ is bounded above and below by a pair of eigenvalues of $\mathsf{L}$. More precisely, applying the Poincar\'e separation theorem to our choice of indices
\begin{equation}
\lambda_m \le \mu_m \le \lambda_{N-r+m}\;\;\;\; m=1,2,\dots,r
\end{equation}
From the expression of the elastic energy in Equation \ref{eq:energy_bilinear} and the  Rayleigh quotients
\begin{equation}
\lambda_m=\frac{
\boldsymbol\phi_m^\mathrm{T}\mathsf{L}\boldsymbol\phi_m
}{
\boldsymbol\phi_m^\mathrm{T}\boldsymbol\phi_m
}
\;\;\;\;
\textrm{ and }\;\;\;\;
\mu_m=\frac{
\boldsymbol\psi_m^\mathrm{T}\mathsf{K}\boldsymbol\psi_m
}{
\boldsymbol\psi_m^\mathrm{T}\boldsymbol\psi_m
}
\end{equation}
we recognize that $\lambda_m/2$ and $\mu_m/2$ are the specific energies per unit displacement (or stiffnesses) of the $m$-th deformation mode in the free and constrained system, respectively.

While in network theory approaches it is common to study the spectrum of $\mathsf{L}$, in boundary value problems such as those relevant to materials mechanics, the quantities of interest depend on the spectrum of $\mathsf{K}$.  We thus define the energy of the $m$-th mode as $\mathcal{E}_m=\mu_m/2$, which serves as our dispersion relation. For every equilibrium equation of the form $\mathsf{K}\mathbf{v}=\mathbf{b}$ such as Equation \ref{eq:solver_problem}, the equilibrium displacement can be expressed as $\mathbf{v}=\mathsf{G}\mathbf{b}$, where the Green operator (discrete Green function) is
$
\mathsf{G}=\sum_{m=1}^{r}
\mu_m^{-1}\boldsymbol\psi_m \boldsymbol\psi_m^\mathrm{T}
$
and the equilibrium displacement has the familiar form
\begin{equation}\label{eq:solution}
\mathbf{v}=\sum_{m=1}^{r}\frac{1}{\mu_m}
\langle \boldsymbol\psi_m,\mathbf{b}\rangle\boldsymbol\psi_m
\end{equation}
where it is expressed as a linear combination of eigenvectors of $\mathsf{K}$, with weights inversely proportional to the corresponding eigenvalues \cite{Moretti2019_EPJB,Greff2024_SciRep}. To elaborate on this concept, we notice that while all eigenpairs $(\mu_m,\boldsymbol\psi_m)$ participate in the solution, each contribution depends also on the scalar product $\langle \boldsymbol\psi_m,\mathbf{b} \rangle$, which measures the compatibility of the deformation mode $\boldsymbol\psi_m$ with the chosen boundary conditions. We refer to low $\mathcal{E}_m$ deformation modes as \textit{soft} deformation modes. Soft modes play a prominent role in Equation \ref{eq:solution} (because of the $1/\mu_m$ factor), and at the same time, they may point to locations with a higher propensity to failure. The density of states
\begin{equation}
\mathcal{D}(\mathcal{E})=\sum_{m=1}^{r}\delta(\mathcal{E}-\mathcal{E}_m)
\end{equation}
measures how prominent modes of energy $\mathcal{E}$ are \textit{globally}. In order to spatially resolve modes of a given energy $\mathcal{E}$, we borrow from quantum mechanics the concept of local density of states \cite{Aizenman2015_Book}, in the form 
\begin{equation}
\mathcal{D}_i(\mathcal{E})=\sum_{m=1}^{r}\delta(\mathcal{E}-\mathcal{E}_m)(\psi_{im})^2,
\end{equation}
where $\psi_{im}$ is the $i$-th component of the $m$-th eigenvector of $\mathsf{K}$. Because of our normalization conventions, summing $\mathcal{D}_i(\mathcal{E})$ over all non-boundary nodes yields $\sum_{i=1}^r\mathcal{D}_i(\mathcal{E})=\mathcal{D}(\mathcal{E})$. We note here that, while forces and energies are locally resolved on edges, deformation modes are locally resolved on nodes, offering a rigorous way of interpolating mechanical observables on nodes.

\section{Simulation results}

\subsection{Fracture strength and localization}
\begin{figure}
\centering
\includegraphics[width=0.8\textwidth]{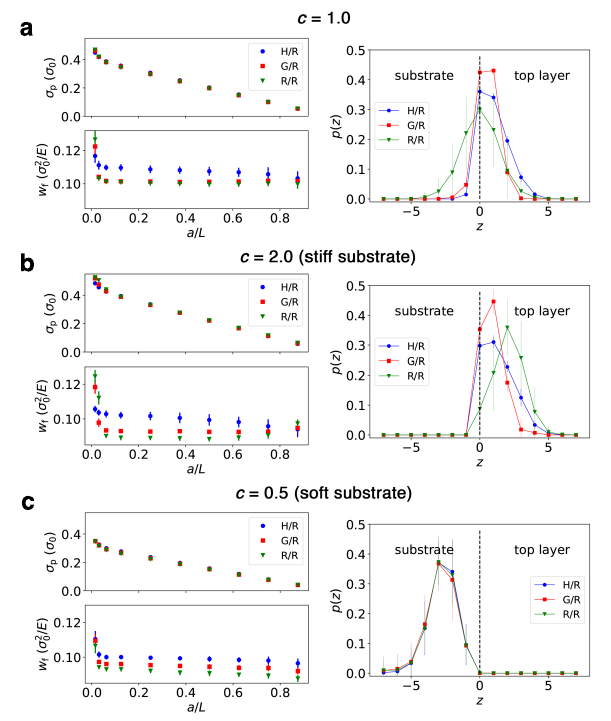}
\caption{Fracture strength and fracture profiles. Data are for systems with $c=1$ (a), $c=2.0$ (b), and $c=0.5$ (c). Peak stress $\sigma_\mathrm{p}$ and specific work of failure $w_\mathrm{f}$ are plotted as functions of notch sizes $a$. The probability $p(z)$ of observing cracks at height $z$ is computed for the un-notched systems ($a=0$).}
\label{fig:fracture}
\end{figure}

We now move to our simulation results. We recall that previous studies have shown that H structures in contact with ideally stiff substrates exhibit enhanced work of failure over reference systems (type R), and have a tendency to localize failure near the interface, allowing for control over fracture surface location. Here, we expect the H/R composite to retain the fracture localization properties observed in simpler systems. However, in our T/S geometry, even the reference R/R system is expected to fail prevalently at $z=0$ for symmetry reasons. We also add the graded G/R system to the comparison, which intuitively should yield similar localization properties as the H/R composite but with a simpler microstructural arrangement. 

Figure \ref{fig:fracture}(a) shows our simulation results for the case where substrate and film have equal stiffness. While minimal differences are observed in the peak stress $\sigma_\mathrm{p}$, the H/R system outperforms both the graded and the reference system in terms of the specific work of failure $w_\mathrm{f}$. Our results are averaged over ensembles of realizations with the same choice of parameters (system type, $c$ and $a$). 
From the final fracture surfaces $z_{\rm f}(x,y)$ we compute the probability $p_{q}(z) = (1/L^2) \sum_{x,y} \delta_{zz_{\rm f}(x,y)}$ that the fracture surface in sample $q$ is, at an arbitrary location parallel to the interface, at distance $z$ from the interface. The ensemble crack height probability  $p(z)$ follows by averaging over all samples. This probability is maximum at or near $z=0$ for all systems, however, it  is close to zero for $z<0$ for the H/R and G/R composites. This indicates that these systems prevent failure in the substrate. While G/R systems behave similar to H/R systems in terms of prescribing failure location, they perform much worse than H/R in terms of work of failure. While the region of reduced $x/y$-edge density that both H and G layers display near the interface plays a rather intuitive role in prescribing failure location, the hierarchical nature of this region in H layers has been shown to suppress crack propagation and facilitate crack arrest \cite{Greff2024_SciRep}, thus inducing a radically different failure mode. H layers attract damage into a softened \textit{buffer region} near $z=0^+$, but they also ensure that the accumulated damage in this region is not conducive to catastrophic failure.

To test the robustness of this picture, we vary the overall stiffness of the S substrate: a stiff substrate ($c=2.0$ in our simulations) intuitively forces failure to occur in the top layer regardless of microstructure, while a soft substrate ($c=0.5$) is expected to do the opposite. We indeed observe these localization properties in Figures \ref{fig:fracture}b,c. However, even for soft substrates, H/R composites still exhibit higher work of failure, as the buffer region still promotes crack arrest.   

\subsection{Energy profiles}

Given that the main differences between the systems studied here reside in the arrangement of $x/y$-edges, which mitigate redistribution of load perpendicular to the loading axis, we ask to which extent these different edge patterns lead to different patterns of elastic stress redistribution. In homogeneous systems exhibiting a crack, transmission of load is maintained by a pattern of load redistribution that concentrates load at the crack tip, thus promoting crack expansion and catastrophic failure. This is in essence what we see in the R/R systems, where crack propagation equates to the breaking of $z$-edges, and stress redistribution implies the loading of $x/y$-edges. However, what happens in H/R and G/R systems? 

\begin{figure}
\centering
\includegraphics[width=0.9\textwidth]{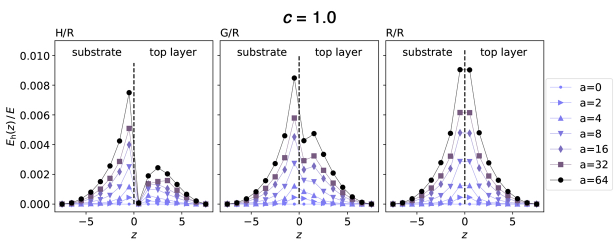}
\caption{Energy profiles. The average fraction $E_\mathrm{h}/E$ of elastic energy stored in $x/y$-edges is plotted against the $z$ coordinate, in case of $c=1$, for different notch sizes $a$, in pristine systems. For $a=0$ $x/y$-edges carry no load, and thus store no energy. For any $a>0$, instead, $x/y$-edges store energy as they participate in the stress redistribution. This role is however suppressed in hierarchical top layers (H/R, for $z>0$).}
\label{fig:energy}
\end{figure}

Let us focus first on the case $c=1$. In Figure \ref{fig:energy} we show the fraction $E_\mathrm{h}/E$ of the elastic energy stored in $x/y$-edges in the pristine state of systems of varying notch sizes, obtained using Equation \ref{eq:energy_restricted}. Henceforth we refer to these plots as \textit{energy profiles}. While energy profiles are trivially symmetric in R/R systems, a clear asymmetry is visible in H/R systems and in G/R systems. In these systems, the preferential removal of $x/y$-edges near the interface has a two-fold consequence. First, both H/R and G/R systems exhibit an elastically softened layer on the T side of the interface. This layer attracts expanding cracks (the elastic energy associated with stress concentrations is reduced) and therefore expanding cracks deflect into this region, as observed in Figure \ref{fig:fracture}. At the same time, in H/R systems, the suppression of stress redistribution ensures that stress in this buffer region does not concentrate at crack tips and incipient cracks are arrested, leading to the observed enhancement in work of failure. Interestingly, G/R systems display a much smaller asymmetry in energy profile. While this is sufficient to make G layers effectively \textit{weak spots}, where failure concentrates, it is not enough to disperse stress concentration at crack tips, and the resulting fracture toughness is not very different from the one in R/R systems. We also notice that no asymmetry arises in the case of $a=0$: in the absence of notches, stress is only carried by the $z$-edges and no energy at all is stored in $x/y$-edges, i.e. no stress redistribution occurs. Even a small notch, e.g. of size $a=2$ is enough to show the differences in response of the three systems. 

\subsection{Deformation modes}

\begin{figure}
\centering
\includegraphics[width=0.75\textwidth]{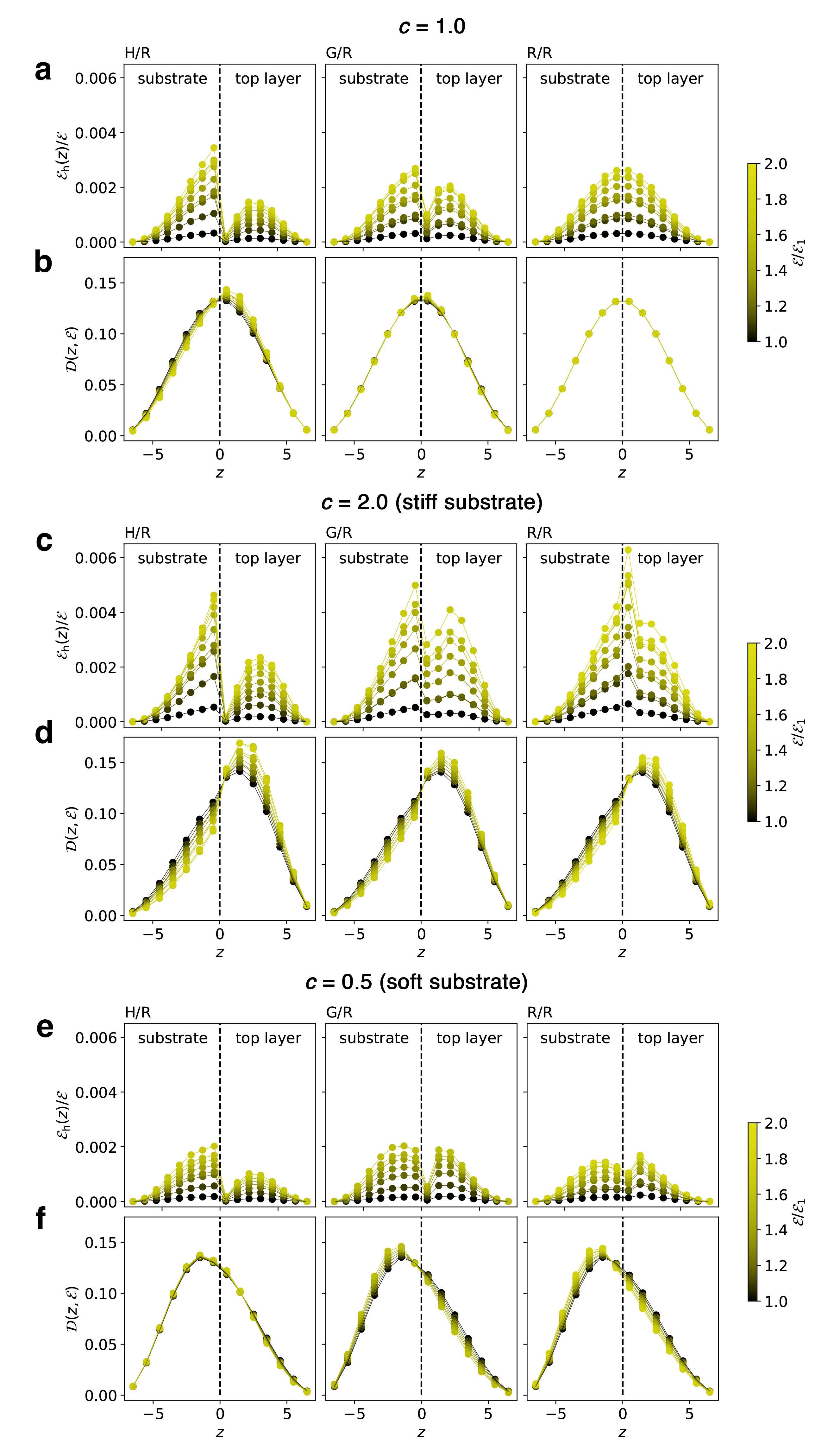}
\caption{Energy profiles and local density of states, for soft deformation modes of varying energy $\mathcal{E}$ near the minimum $\mathcal{E}_1$, in systems with $c=1.0$ (a,b), $c=2.0$ (c,d), and $c=0.5$ (e,f). The interval of normalized energies $1.0\le \mathcal{E}/\mathcal{E}_1\le 2.0$ is divided into 10 bins. Darker colors indicate bins of lower normalized energy. $\mathcal{D}(z,\mathcal{E})$ is measued in units of $\kappa_0^{-1}$.}
\label{fig:allspectra}
\end{figure}

To further investigate the asymmetry in energy profiles, we conduct the same analysis as above, this time for all deformation modes with a given energy $\mathcal{E}$. To do so, we resort to our spectral formalism. We consider the range $1.0\le \mathcal{E}/\mathcal{E}_1\le 2.0$ of normalized energies of soft deformation modes near the minimum energy $\mathcal{E}_1$, and partition it into $10$ bins of width $0.1$. In Figure \ref{fig:allspectra}a we show the average energy profiles computed for the modes within each energy bin. In Figure \ref{fig:allspectra}b we show the average local density $\mathcal{D}(z,\mathcal{E})$ of states, averaged over nodes at height $z$ for the same energy bins as in Figure \ref{fig:allspectra}a. For these plots, we again consider a fixed notch length of $a=2$.  

Energy profiles mimic those encountered in our simulations (Figure \ref{fig:energy}), H layers always develop buffer regions where energy accumulation in $x/y$-edges is reduced. These findings are mirrored by the local density of states of soft modes, which is higher in layers where energy profiles are lower than those of our reference systems. The enhanced local density of soft modes in the H phase points precisely to the buffer region hypothesized above. 
As expected, a similar phenomenology is encountered in G layers, but in much weaker form.

We can now test the robustness of these results by varying our stiffness parameter $c$. Figures \ref{fig:allspectra}c-d and \ref{fig:allspectra}e-f show energy profiles and local density of states in the cases $c=2.0$ and $c=0.5$ respectively. Energy profiles are obviously affected in both cases, but a much clearer picture is provided by the local density of states. In particular, while the local density of states is now skewed towards the softer layers, H/R systems always deviate from the reference R/R behavior, exhibiting higher-than-reference values in correspondence with the buffer region.

Figure \ref{fig:allspectra} allows us to detect asymmetry in energy profiles and local densities of modes of extremely low energy. We note here that the lowest and highest eigenvalues for all systems are approximately $\mu_1\approx 0.044$, $\mu_r\approx 11$ (eigenvalues are measured in units of $\kappa_0$). Keeping in mind that eigenmode energies are $\mathcal{E}_m=\mu_m/2$, Figure  \ref{fig:allspectra} barely covers the lower spectral edge, i.e. the range of eigenvalues that are closest to $\mu_1$ (80 of them, in the range considered in Figure \ref{fig:allspectra}). While the $1/\mu_m$ factor in Equation \ref{eq:solution} suggests that modes of lower $\mu_{\rm m}$ may have greater weight in the eigenvector expansion of the equilibrium solution $\mathbf{v}$, the scalar product $\langle \boldsymbol{\psi}_m,\mathbf{b}\rangle$ may play an equally important role. Are modes of higher energy relevant to the problem at hand and, if so, do we still observe asymmetric energy profiles for those modes?

In order to answer these questions, we first compute the spectral decomposition of the energy of our system for the equilibrium problem with unit displacement applied to the top boundary. To this end we plug the expressions for $\mathsf{L}$ and $\mathbf{u}$ from Equation \ref{eq:blocks} into Equation \ref{eq:energy_bilinear} and obtain
\begin{equation}
E=\frac{1}{2}\left(
\langle \mathbf{v},\mathsf{K}\mathbf{v}  \rangle
+2\langle \mathbf{v},\mathsf{R}\mathbf{w} \rangle
+\langle \mathbf{w},\mathsf{S}\mathbf{w}  \rangle
\right),
\end{equation}
where we recall that $\mathbf{v}$ and $\mathbf{w}$ are the displacements at non-boundary and boundary nodes, respectively. At equilibrium and in the absence of body forces $\mathsf{K}\mathbf{v}+\mathsf{R}\mathbf{w}=0$, yielding
\begin{equation}
E=\frac{1}{2}\left(
\langle \mathbf{v},\mathsf{R}\mathbf{w} \rangle
+\langle \mathbf{w},\mathsf{S}\mathbf{w}  \rangle
\right).
\end{equation}
Recalling that $\mathbf{b}=-\mathsf{R}\mathbf{w}$ and expressing $\mathbf{v}$ via its eigenvector decomposition we finally obtain
\begin{equation}
E=\frac{1}{2}\left(
\langle \mathbf{w},\mathsf{S}\mathbf{w}  \rangle - \sum_{m=1}^r \zeta_m
\right)
\end{equation}
where the weight 
\begin{equation}
\zeta_m =  \frac{1}{\mu_m}\langle\boldsymbol{\psi}_m,\mathbf{b} \rangle^2
\end{equation}
is a measure of the contribution of the $m-$th mode to the total energy. 

\begin{figure}
\centering
\includegraphics[width=0.8\textwidth]{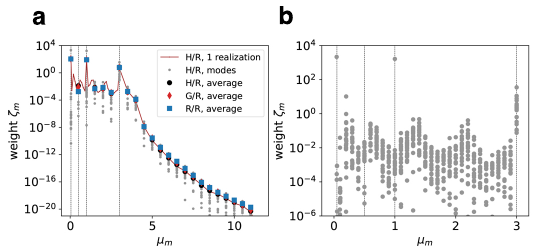}
\caption{Weights of individual modes in the computation of the elastic energy at equilibrium, with $c=1$. (a) Full eigenvalue range. Small grey dots: individual modes, sampled as groups of 20, for a single realization of the H/R system. Thin red line, average of the single realization H/R data, with a denser sampling in the region with $\mu_{\rm m}\le3$. Large symbols: averages over multiple realizations, for the H/R, G/R and R/R systems. (b) Close-up of the data for a single H/R realization in the $\mu_{\rm m}\le3$ region, this time  for a denser sampling. In both panels, vertical dotted lines indicate the values $\mu_{\rm m}$ that are later considered in Figure \ref{fig:spectra_highmu}.}
\label{fig:coefficients}
\end{figure}

\begin{figure}
\centering
\includegraphics[width=0.9\textwidth]{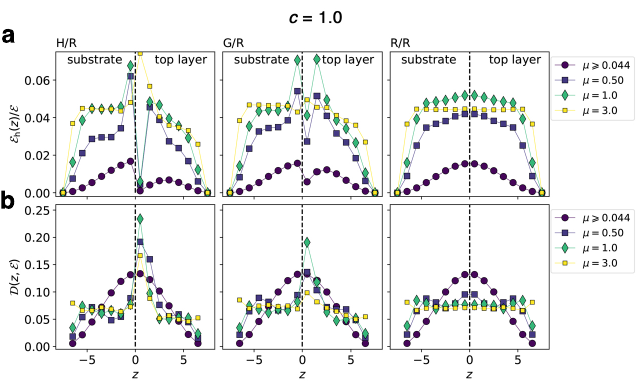}
\caption{Energy profiles and local density of states, for soft deformation modes in the extended energy range with $\mu\le 3.0$ ($\mathcal{E}\le 1.5$). Modes are sampled in groups of 20, at the lowest spectral edge (indicated as $\mu\ge0.044$) and around values 0.50, 1.0 and 3.0 (the same values are highlighted by vertical dotted lines in Figure \ref{fig:coefficients}).  $\mathcal{D}(z,\mathcal{E})$ is measued in units of $\kappa_0^{-1}$.}
\label{fig:spectra_highmu}
\end{figure}

In the following, we focus on systems with $c=1.0$. Figure \ref{fig:coefficients}a shows the values of $\zeta_m$, as sampled across the whole eigenvalue range, using the following procedure. Groups of 20 adjacent eigenvalues, along with their respective eigenvectors, are sampled at the lower spectral edge and at regular intervals $\mu_{\rm m}=0.5,\,1.0,\,1.5\dots 11.0$. The corresponding values of $\zeta_m$ are computed (small grey dots) and averaged across multiple realizations, for our three systems (large symbols). Figure \ref{fig:coefficients}b shows a close-up of the $\mu\le 3$ range of higher weights, where a denser sampling is considered, for a single realization of the H/R system. While the highest weights are indeed found near the lower spectral edge $\mu_{\rm m}\ge 0.044$, the range of non-negligible weights extends up to $\mu_{\rm m}\approx 3$, suggesting that while soft modes indeed contribute more significantly to the energy of the system, what we should call soft modes extends well beyond the lower spectral edge and has a clear upper bound. Beyond $\mu_{\rm m}=3$ the effect of the factor $1/\mu_{\rm m}$ becomes more prominent and weights become negligible. Now that we have identified an eigenvalue range of interest, we can look at the energy profiles, for the deformation modes in that range. Figure \ref{fig:spectra_highmu} shows the energy profiles and local density of states for the groups of modes at the lower spectral edge and at $\mu_{\rm m}\approx 0.5,\,1.0,\,3.0$ (these same groups are highlighted in Figure \ref{fig:coefficients} with vertical dotted lines). H/R systems still exhibit asymmetric energy profiles and in particular store very low values of elastic energy in the horizontal edges at $z=1/2$. This mirrors the behavior at very low energies ($\mu_{\rm m}\ge 0.044$ and Figure \ref{fig:allspectra}). The only exception is found for the highest $\mu_{\rm m}$ in the range (small yellow squares), where energy profiles are still asymmetric and a peak appears at $z=1/2$. The weights of these exceptional configurations are however between one and four orders of magnitude less than those of the dominant modes at lower $\mu_{\rm m}$ (Figure \ref{fig:coefficients}).

Figure \ref{fig:spectra_highmu}b shows that, in the here investigated extended energy range, the local density of states for the H/R systems has an even more pronounced peak as compared to the subtle asymmetry encountered in Figure \ref{fig:allspectra} for low energy modes. This confirms that the properties of soft modes near the lower spectral edge carry over to higher energies and are representative of the majority of modes that are relevant to the equilibirum properties of the system.

Finally, we note that our results on energy profiles and local densities of states for ranges of deformation modes in Figures \ref{fig:allspectra} and \ref{fig:spectra_highmu} are more general than the specific equilibrium problem that we study here. Given the Laplacian $\mathsf{L}$ of a free problem, what fixes the stiffness matrix $\mathsf{K}$, and thus the deformation modes $\boldsymbol{\psi}_m$ of the constrained problem, is the choice of boundary nodes. The actual boundary values and possible body forces do not affect the individual deformation modes -- they only affect their contributions, e.g. via the $\zeta_m$ weights.

\section{Discussion and conclusions}
Our results show that under rather general assumptions, graded structures can be employed to prescribe failure location, by imposing density gradients along the loading direction. The question whether these systems are also more flaw tolerant is more complex. Density gradients alone do not yield higher work of failure, and more complex microstructures that are both graded and conducive to some form of crack arrest are necessary, such as the hierarchical layers considered here. Interestingly, altering the stiffness/hardness of the substrate can override the failure localization properties of a graded top layer (a soft-enough substrate always breaks by crack propagation within the substrate, as for $c=0.5$), but even under these extreme conditions hierarchical systems still exhibit higher work of failure. Importantly, none of the structures considered here display significant differences in peak load: our results are significant to the resilience of these systems, rather than to their peak strength.

The enhanced fracture toughness of hierarchical layers has been recently linked to the inherent anisotropy of the elastic Green function of these systems, which is produced by their distinctive  lamellar cut patterns \cite{Greff2024_SciRep}. With our current results, we can elaborate on that picture, observing that the enhanced resilience of hierarchical systems relies on their ability to create a buffer region where stress redistribution and energy storage are suppressed in the load-perpendicular direction. As a consequence, significant damage is stored, but also rendered unable to propagate. Tuning the location of this damage reservoir, one can prescribe where failure is going to concentrate. 

Our network model allows us to clearly differentiate between load carrying elements ($z$-edges) and crosslinking elements ($x/y$-edges), providing us with the correct tools to resolve quantities locally. This is especially important in  network-like microstructures, where energies are stored in edges (trusses, beams, fibers). Edges of different orientations play different roles, and node-resolved energy densities are not easily defined. 

More in general, any network material model of this type will be edge-centric in nature, as its constitutive laws are defined on edges. The material is essentially made of edges, not nodes. This is different from what is found in most network models of techno-social systems, where nodes are the agents in the system and node-resolved properties matter, such as connectivity, clustering, modularity and centrality \cite{Barabasi1999_Scaling,PastorSatorras2015_RMP,Newman2018_Book}. Our formalism here aims at establishing a basic network vocabulary for materials, at least in the context of elasticity and brittle failure. We chose our formalism to be as close as possible to that in use in discrete differential geometry applications of relevance to continuum mechanics, electromagnetism and geometry processing \cite{Bossavit2001_GFD,Hiptmair2001_algebraic,Hirani2003_Thesis,Desbrun2008_book,Grady2010_Book,Arnold2018_Book} hopefully paving the way for future generalizations of our results to continua, and for the application of concepts of Riemannian geometry to materials problems. In this respect, we notice that concepts of vector flow on smooth manifolds already have applications in hydrodynamics \cite{Arnold2014,Chen2023_Fluids,Zhu2025_PRSA}, and appear especially promising for visualizing patterns of load redistribution in materials and metamaterials.  

Finally, our spectral formalism allows us to introduce a physically meaningful property that is defined on nodes, the local density of states, which resolves collective deformation modes of prescribed energy on node locations. The use of the concept of local density of states in complex network models is not new \cite{Margiotta2019_JSTAT,daSilva2025_SciPost}. Here we have shown its relevance in microstructure mechanics, where soft modes are the leading terms in the expansion of equilibrium problems, and may be used to visualize load and damage patterns. We emphasize that this methodology carries over to any linear system described by a stiffness matrix like our $\mathsf{K}$, as found for instance in beam network models, as well as in finite element analysis. More in general, in systems where complex microstructural arrangements yield spatially heterogeneous response, local spectral measures may prove useful in interpreting digital image correlation (DIC)  data \cite{Mustalahti2010,Miksic2011}. Similarly, local spectral measures may provide input for data-driven materials design approaches in problems of failure forecast \cite{hiemer2022predicting}, and strength optimization \cite{yu2022multiresolution,zaiser2023disordered,luu2023,buehler2023msm}.


\funding{Funded by the Deutsche Forschungsgemeinschaft (DFG, German Research Foundation) – Project numbers 377472739/GRK 2423/2-2023 and 394689530.
}




\bibliography{adhesion3}

\end{document}